\documentclass[12pt]{article} 
\usepackage{amsmath,amssymb,mathrsfs,bbm,bm}
\usepackage{natbib}
\usepackage[usenames, dvipsnames]{color}
\usepackage[colorlinks=true, citecolor=blue, urlcolor=blue, linkcolor=blue]{hyperref}
\usepackage{setspace}
\usepackage[lofdepth,lotdepth]{subfig}
\usepackage{booktabs} 
\usepackage{fancyvrb}
\usepackage[framemethod=TikZ]{mdframed}
\usepackage[space]{grffile}
\usepackage[nottoc]{tocbibind}
\usepackage{tcolorbox}

\usepackage{float}

\usepackage[margin = 1in]{geometry}

\numberwithin{figure}{section}
\numberwithin{table}{section}
\numberwithin{equation}{section}

\DeclareMathOperator*{\argmin}{arg\,min}

\newcommand{\E}{\mathbb{E}}
\renewcommand{\P}{\mathbb{P}}

\newcommand{\I}{\mathbbm{1}}

\begin{document}

\onehalfspacing

\begin{titlepage}
  \title{\bf The Regression Discontinuity Design\footnote{We thank Rich Nielsen for his comments and suggestions on a previous version of this chapter.}}
  \author{
    Matias D. Cattaneo\thanks{Department of Operations Research and Financial Engineering, Princeton University.}
    \and Roc\'{i}o Titiunik\thanks{Department of Politics, Princeton University.}
    \and Gonzalo Vazquez-Bare\thanks{Department of Economics, University of California at Santa Barbara.}
}

\date{\large{\vspace{0.5in} \today}}
\end{titlepage}
\maketitle

\thispagestyle{empty}

\begin{center}
	Handbook chapter published in\medskip\\
	\textit{Handbook of Research Methods in Political Science and International Relations}\medskip\\
	Sage Publications, Ch. 44, pp. 835-857, June 2020. \bigskip\\
	Published version:\medskip\\
	\url{http://dx.doi.org/10.4135/9781526486387.n47}
	
\end{center}
\thispagestyle{empty}

\newpage\pagenumbering{roman}\setcounter{page}{1}
{\singlespacing\tableofcontents}

\newpage 

\newpage\pagenumbering{arabic}\setcounter{page}{1}

\section{Introduction}\label{sec:intro}

The Regression Discontinuity (RD) design has emerged in the last decades as one of the most credible non-experimental research strategies to study causal treatment effects. The distinctive feature behind the RD design is that all units receive a score, and a treatment is offered to all units whose score exceeds a known cutoff, and withheld from all the units whose score is below the cutoff. Under the assumption that the units' characteristics do not change abruptly at the cutoff, the change in treatment status induced by the discontinuous treatment assignment rule can be used to study different causal treatment effects on outcomes of interest. 

The RD design was originally proposed by \citet*{Thistlethwaite-Campbell_1960_JEP} in the context of an education policy, where an honorary certificate was given to students with test scores above a threshold. Over time, the design has become common in areas beyond education, and is now routinely used by scholars and policy-makers across the social, behavioral, and biomedical sciences. In particular, the RD design is now part of the standard quantitative toolkit of political science research, and has been used to study the effect of many different interventions including party incumbency, foreign aid, and campaign persuasion.

In this chapter, we provide an overview of the basic RD framework, discussing the main assumptions required for identification, estimation, and inference. We first discuss the most common approach for RD analysis, the \textit{continuity-based} framework, which relies on assumptions of continuity of the conditional expectations of potential outcomes given the score, and defines the basic parameter of interest as an average treatment effect at the cutoff. We discuss how to estimate this effect using local polynomials, devoting special attention to the role of the bandwidth, which determines the neighborhood around the cutoff where the analysis is implemented. We consider the bias-variance trade-off inherent in the most common bandwidth selection method (which is based on mean-squared-error minimization), and how to make valid inferences with this bandwidth choice. We also discuss the local nature of the RD parameter, including recent developments in extrapolation methods that may enhance the external validity of RD-based results. 

In the second part of this chapter, we overview an alternative framework for RD analysis that, instead of relying on continuity of the potential outcome regression functions, makes the assumption that the treatment is as-if randomly assigned in a neighborhood around the cutoff. This interpretation was the intuition provided by \cite{Thistlethwaite-Campbell_1960_JEP} in their original contribution, though it now has become less common due to the stronger nature of the assumptions it requires. We discuss situations in which this \textit{local randomization} framework for RD analysis may be relevant, focusing on cases where the running variable has mass points, which occurs very frequently in applications. 

To conclude, we discuss a battery of data-driven falsification tests that can provide empirical evidence about the validity of the design and the plausibility of its key identifying assumptions. These falsification tests are intuitive and easy to implement, and thus should be included as part of any RD analysis in order to enhance its credibility and replicability.

Due to space limitations, we do not discuss variations and extensions of the canonical (sharp) RD designs, such as fuzzy, kink, geographic, multi-cutoff or multi-score RD designs. A practical introduction to those topics can be found in \citet*{Cattaneo-Idrobo-Titiunik_2019_Book1,Cattaneo-Idrobo-Titiunik_2019_Book2}, in the recent edited volume \citet*{Cattaneo-Escanciano_2017_AIE}, and the references therein. For a recent review on program evaluation methods see \citet*{Abadie-Cattaneo_2018_ARE}.

\section{General Setup}\label{sec:setup}
We start by introducing the basic notation and framework. We consider a study where there are multiple units from a population of interest (such as politicians, parties, students, households or firms), and each unit $i$ has a \textit{score} or \textit{running variable}, denoted by $X_i$. This running variable could be, for example, a party's vote share in a congressional district, a student's score from a standardized test, a household's poverty index, or a firm's total revenues in a certain period of time. This running variable may be continuous, in which case no two units will have the same value of $X_i$, or not, in which case the same value of $X_i$ might be shared by multiple units. The latter case is usually called ``discrete'', but in many empirical applications the score variable is actually both.

In the simplest RD design, each unit receives a binary treatment $D_i$ when their score exceeds some fixed threshold $c$, and does not receive the treatment otherwise. This type of RD design is commonly known as the \textit{sharp RD design}, where the word \textit{sharp} refers to the fact that the assignment of treatment coincides with the actual treatment taken---that is, compliance with treatment assignment is perfect. When treatment compliance is imperfect, the RD design becomes a \textit{fuzzy RD design} and its analysis requires additional methods beyond the scope of this chapter (see the Introduction for references). The methods described here for analyzing sharp RD designs can be applied directly in the context of fuzzy RD designs when the parameter of interest is the intention-to-treat effect.

The sharp RD treatment assignment rule can be formally written as
\begin{equation}\label{eq:sharp}
D_i=\I(X_i\ge c)=
\begin{cases}
1 & \quad \text{if }X_i\ge c \\
0 & \quad \text{if }X_i< c 
\end{cases},
\end{equation}
where $\I(\cdot)$ is the indicator function. For example, $D_i$ could be a scholarship for college students that is assigned to those with a score of 7 or higher in an entry exam on a scale from 0 to 10. In this example, $X_i$ is the exam score, $c=7$ is the cutoff used for treatment assignment, and $D_i=\I(X_i\ge 7$) is the binary variable that indicates receipt of the scholarship.

Our goal is to assess the effect of the binary treatment $D_i$ on a certain outcome variable. For instance, in the previous scholarship example, we may be interested in analyzing whether the scholarship increases the academic performance during college or the probability of graduating. This problem can be formalized within the potential outcomes framework \citep{Imbens-Rubin_2015_Book}. In this framework, each unit $i$ from the population of interest has two potential outcomes, denoted $Y_i(1)$ and $Y_i(0)$, which measure the outcome that would be observed for unit $i$ with and without treatment, respectively. For example, for a certain college student $i$, $Y_i(1)$ could be the student's GPA at a certain stage had the student received the scholarship, and $Y_i(0)$ the student's GPA had she not received the scholarship. The individual-level treatment ``effect'' for unit $i$ is defined as the difference between the potential oucomes under treatment and control status, $\tau_i=Y_i(1)-Y_i(0)$.

Because the same unit can never be observed under both treated and control status (a student can either receive or not receive the scholarship, but not both), one of the potential outcomes is always unobservable. The observed outcome, denoted $Y_i$, equals $Y_i(1)$ when $i$ is treated and $Y_i(0)$ if $i$ is untreated, that is,
\begin{equation}\label{eq:obsout}
Y_i=Y_i(1)\cdot D_i+Y_i(0)\cdot (1-D_i)=\begin{cases}
Y_i(1) &\quad \text{if }D_i=1 \\
Y_i(0) &\quad \text{if }D_i=0
\end{cases}.
\end{equation}
The observed outcome can never provide information on both potential outcomes. Hence, for each unit in the population, one of the potential outcomes is observed, and the other one is a \textit{counterfactual}. This problem is known as the \textit{fundamental problem of causal inference} \citep{Holland_1986}. 

The RD design provides a way to address this problem by comparing treated units that are ``slightly above'' the cutoff to control units that are ``slightly below'' it. The rationale behind this comparison is that, under appropriate assumptions that will be made more precise in the upcoming sections, treated and control units in a small neighborhood or \textit{window} around the cutoff are comparable in the sense of having similar observed and unobserved characteristics (with the only exception being treatment status). Thus, observing the outcomes of units just below the cutoff provides a valid measure of the average outcome that treated units just above the cutoff would have had if they had not received the treatment. 

In the remainder of this chapter, we describe two alternative approaches for analyzing RD designs. The first one, which we call the \textit{continuity-based framework}, assumes that the observed sample is a random draw from an infinite population of interest, and invokes assumptions of continuity. In this framework, identification of the parameter of interest, defined precisely in the next section, relies on assuming that the average potential outcomes given the score are continuous as a function of the score. This assumption implies that the researcher can compare units marginally above the cutoff to units marginally below to identify (and estimate) the average treatment effect at the cutoff.

The second approach for RD analysis, which we call the \textit{local randomization framework}, assumes that the treatment of interest is as-if randomly assigned in a small region around the cutoff. This approach formalizes the interpretation of RD designs as local experiments, and allows researchers to use the standard tools from the classical analysis of experiments. In addition, if the researcher is willing to assume that potential outcomes are fixed (non-random) and that the $n$ units that are observed in the sample conform the finite population of interest, this approach also allows the researcher to use finite-sample exact randomization inference tools, which are specially appealing in applications where the number of observations near the cutoff is small.

For both frameworks, we discuss the parameters of interest, estimation, inference, and bandwidth or window selection methods. We then compare the two approaches and provide a series of falsification methods that are commonly employed to assess the validity of the RD design. See also \citet*{Cattaneo-Titiunik-VazquezBare_2017_JPAM} for an overview and practical comparisons between these RD approaches.

\section{The Continuity-Based Framework}\label{sec:continuity}

Under the continuity-based framework, the observed data $\{Y_i(1),Y_i(0),X_i,D_i\}$, for $i=1,2,\ldots,n$, is a random sample from an infinite population of interest (or data generating process). The main objects of interest under this framework are the conditional expectation functions of the potential outcomes,
\begin{equation}\label{eq:cef}
\mu_1(x)=\E[Y_i(1)|X_i=x]\qquad\text{and}\qquad \mu_0(x)=\E[Y_i(0)|X_i=x],
\end{equation}
which capture the population average of the potential outcomes for each value of the score. In the sharp RD design, for each value of $x$, only one of these functions is observed: $\mu_1(x)$ is observed for $x$ at or above the cutoff, and $\mu_0(x)$ is observed for values of $x$ below the cutoff. 

The observed conditional expectation function is
\begin{equation}\label{eq:obscef}
\mu(x)=\E[Y_i|X_i=x]=\begin{cases}
\mu_1(x) &\quad \text{if }x \ge c \\
\mu_0(x) &\quad \text{if }x<c  \text{.}
\end{cases}
\end{equation}

We start by defining the function $\tau(x)$, which gives the average treatment effect conditional on $X_i=x$: 
\begin{equation}\label{eq:te}
\tau(x)=\E[Y_i(1)-Y_i(0)|X_i=x]=\mu_1(x)-\mu_0(x)\text{.}
\end{equation}
The first step is to establish conditions for identification, that is, conditions under which we can write the parameter of interest, which depends on unobservable quantities due to the fundamental problem of causal inference, in terms of observable (i.e., identifiable) and thus estimable quantities. In the continuity-based framework, the key assumption for identification is that $\mu_1(x)$ and $\mu_0(x)$ are continuous functions of the score at the cutoff point $x=c$. Intuitively and informally, this condition states that the observable and unobservable characteristics that determine the average potential outcomes do not jump abruptly at the cutoff. When this assumption holds, the only difference between units on opposite sides of the cutoff whose scores are ``very close'' to the cutoff is their treatment status.

Intuitively, we can think that treated and control units with very different score values will generally be very different in terms of important observable and unobservable characteristics affecting the outcome of interest but, as their scores approach the cutoff and become similar in that dimension, the only remaining difference between them will be their treatment status, thus ensuring comparability between units just above and just below the cutoff, at least in terms of their potential outcome mean regression functions.

More formally, \citet*{Hahn-Todd-vanderKlaauw_2001_ECMA} showed that, when conditional expectation functions are continuous in $x$ at the cutoff level $x=c$,
\begin{equation}\label{eq:sharp_id}
\tau(c) = \lim_{x\downarrow c}\E[Y_i|X_i=x]-\lim_{x\uparrow c}\E[Y_i|X_i=x],
\end{equation}
so that the difference between average observed outcomes for units just above and just below the cutoff is equal to the average treatment effect at the cutoff, $\tau(c)=\E[Y_i(1)-Y_i(0)|X_i=c]$. Note that this identification result expresses the estimand $\tau(c)$, which is unobservable, as a function of two limits that depend only on observable (i.e., identifiable) quantities that are estimable from the data.

As a consequence, in a sharp RD design, a natural parameter of interest is $\tau(c)$, the average treatment effect at the cutoff. This parameter captures the average effect of the treatment on the outcome of interest, given that the value of the score is equal to the cutoff. It is useful to compare this parameter to the average treatment effect, $\mathtt{ATE} = \E[Y_i(1)-Y_i(0)]$, which is the difference that we would see in average outcomes if all units were switched from control to treatment. In contrast to $\mathtt{ATE}$, which is a weighted average of $\tau(x)$ over $x$ because $\mathtt{ATE}=\E[\tau(X_i)]$, $\tau(c)$ is only the average effect of the treatment at a particular value of the score, $x=c$. For this reason, the RD parameter of interest  $\tau(c)$ is often referred to as a \textit{local} average treatment effect, because it is only informative of the effect of the treatment for units whose value of the score is at (or, loosely speaking, in a local neighborhood of) the cutoff. This limits the external validity of the RD parameter $\tau(c)$. A recent and growing literature studies how to extrapolate treatment effects in RD designs \citep*{Angrist-Rokkanen_2015_JASA,Dong-Lewbel_2015_ReStat,Cattaneo-Keele-Titiunik-VazquezBare_2016_JOP,Bertanha-Imbens_2019_JBES,Cattaneo-Keele-Titiunik-VazquezBare_2019_wp}.

The main advantage of the identification result in \eqref{eq:sharp_id} is that it relies on continuity conditions of $\mu_1(x)$ and $\mu_0(x)$ at $x=c$, which are nonparametric in nature and reasonable in a wide array of empirical applications. Section \ref{sec:falsification} describes several falsification strategies to provide indirect empirical evidence to assess the plausibility of this assumption. Assuming continuity holds, the estimation of the RD parameter  $\tau(c)$  can proceed without making parametric assumptions about the particular form of $\E[Y_i|X_i=x]$. Instead, estimation can proceed by using nonparametric methods to approximate the regression function $\E[Y_i|X_i=x]$, separately for values of $x$ above and below the cutoff.

However, estimation and inference via nonparametric local approximations near the cutoff is not without challenges. When the score is continuous, there are in general no units with value of the score exactly equal to the cutoff. Thus, estimation of the limits of  $\E[Y_i|X_i=x]$ as $x$ tends to the cutoff from above or below will necessarily require extrapolation. To this end, estimation in RD designs requires specifying a neighborhood or bandwidth around the cutoff in which to approximate the regression function $\E[Y_i|X_i=x]$, and then, based on that approximation, calculate the value that the function has exactly at $x=c$. In what follows, we describe different methods for estimation and bandwidth selection under the continuity-based framework.

\subsection{Bandwidth Selection}

Selecting the bandwidth around the cutoff in which to estimate the effect is a crucial step in RD analysis, as the results and conclusions are typically sensitive to this choice. We now briefly outline some common methods for bandwidth selection in RD designs. See also \citet*{Cattaneo-VazquezBare_2016_ObsStud} for an overview of neighborhood selection methods in RD designs.

The approach for bandwidth selection used in early RD studies is what we call \textit{ad-hoc} bandwidth selection, in which the researcher chooses a bandwidth without a systematic data-driven criterion, perhaps relying on intuition or prior knowledge about the particular context. This approach is not recommended since it lacks objectivity, does not have a rigorous justification and, by leaving bandwidth selection to the discretion of the researcher, opens the door for specification searches. For these reasons, the ad-hoc approach to bandwidth selection has been replaced by systematic, data-driven criteria.

In the RD continuity-based framework, the most widely used bandwidth selection criterion in empirical practice is the \textit{mean squared error} (MSE) criterion \citep*{Imbens-Kalyanaraman_2012_REStud,Calonico-Cattaneo-Titiunik_2014_ECMA,Arai-Ichimura_2018_QE,Calonico-Cattaneo-Farrell-Titiunik_2019_RESTAT}, which relies on a tradeoff between the bias and variance of the RD point estimator. The bandwidth determines the neighborhood of observations around the cutoff that will be used to approximate the unknown function $\E[Y_i|X_i=x]$ above and below the cutoff. Intuitively, choosing a very small bandwidth around the cutoff will tend to reduce the misspecification error in the approximation, thus reducing bias. A very small bandwidth, however, requires discarding a large fraction of the observations and hence reduces the sample, leading to estimators with larger variance. Conversely, choosing a very large bandwidth allows the researcher to gain precision using more observations for estimation and inference, but at the expense of a larger misspecification error, since the function $\E[Y_i|X_i=x]$   now has to be approximated over a larger range. The goal of bandwidth selection methods based on this tradeoff is therefore to find the bandwidth that optimally balances bias and variance.

We let $\hat{\tau}$ denote a local polynomial estimator of the RD treatment effect $\tau(c)$---we explain how to construct this estimator in the next section. For a given bandwidth $h$ and a total sample size $n$, the MSE of $\hat{\tau}$ is
\begin{equation}\label{eq:mse}
\mathsf{MSE}(\hat{\tau})=\mathsf{Bias}^2(\hat{\tau})+\mathsf{Variance}(\hat{\tau}) = \mathscr{B}^2 + \mathscr{V}\text{,}
\end{equation}
which is the sum of the squared bias and the variance of the estimator. The MSE-optimal bandwidth, $h_{\mathsf{MSE}}$, is the value of $h$ that balances bias and variance by minimizing the MSE of $\hat{\tau}$, 
\begin{equation}\label{eq:hmse}
h_{\mathsf{MSE}}=\argmin_{h>0} \mathsf{MSE}(\hat{\tau}) \text{.}
\end{equation}

The shape of the MSE depends on the specific estimator chosen. For example, when $\hat{\tau}$ is obtained using local linear regression (LLR), which will be discussed in the next section, the MSE can be approximated by
\[\mathsf{MSE}(\hat{\tau})\approx h^4\mathsf{B}^2+\frac{1}{nh}\mathsf{V}\]
where $\mathsf{B}$ and $\mathsf{V}$ are constants that depend on the data generating process and specific features of the estimator used. This expression clearly highlights how a smaller bandwidth reduces the bias term while increasing the variance and vice versa. In this case, the optimal bandwidth, simply obtained by setting the derivative of the above expression with respect to $h$ equal to zero, is
\begin{equation}\label{eq:hmse_llr}
h_{\mathsf{MSE}}^{\mathsf{LLR}}=\mathsf{C}_\mathsf{MSE} \cdot n^{-1/5},
\end{equation}
where the constant $\mathsf{C}_\mathsf{MSE}=(\mathsf{V}/(4\mathsf{B}^2))^{1/5}$ is unknown but estimable. This shows that the MSE-optimal bandwidth for a local linear estimator is proportional to $n^{-1/5}$.

While $h_{\mathsf{MSE}}$ is optimal for point estimation, it is generally not optimal for conducting inference. \citet*{Calonico-Cattaneo-Farrell_2018_JASA,Calonico-Cattaneo-Farrell_2019_CEoptimal,Calonico-Cattaneo-Farrell_2019_CERD} show how to choose the bandwidth to obtain confidence intervals minimizing the coverage error probability (CER). More precisely, let $\mathsf{IC}(\hat{\tau})$ be an $\alpha$-level confidence interval for the RD parameter $\tau(c)$ based on the estimator $\hat{\tau}$. A CER-optimal bandwidth makes the coverage probability as close as possible to the desired level $1-\alpha$:
\begin{equation}\label{eq:hcer}
h_{\mathsf{CER}}=\argmin_{h>0} |\P[\tau(c)\in \mathsf{IC}(\hat{\tau})]-(1-\alpha)| \text{.}
\end{equation}
For the case of local linear regression, the CER-optimal $h$ is
\[h_\mathsf{CER}^{\mathsf{LLR}}=\mathsf{C}_\mathsf{CER}\cdot n^{-1/4},\]
where again the constant $\mathsf{C}_\mathsf{CER}$ unknown, because depends in part on the data generating process, but estimable. Hence, the CER-optimal bandwidth is smaller than the MSE-optimal bandwidth, at least in large samples.

Based on the ideas above, several variations of optimal bandwidth selectors exist, including one-sided CER-optimal and MSE-optimal bandwidths with and without accounting for covariate adjustment, clustering, or other specific features. In all cases, these bandwidth selectors are implemented in two steps: first the constant (e.g., $\mathsf{C}_\mathsf{MSE}$ or $\mathsf{C}_\mathsf{CER}$) is estimated, and then the bandwidth is chosen using that preliminary estimate and the appropriate rate formula (e.g., $n^{-1/5}$ or $n^{-1/4}$).

\subsection{Estimation and Inference}\label{sec:estinf1}

Given a bandwidth $h$, continuity-based estimation in RD designs consists on estimating the outcome regression functions, given the score, separately for treated and control units whose scores are within the bandwidth. Recall from Equation (\ref{eq:sharp_id}) that we need to estimate the limits of the conditional expectation function of the observed outcome from the right and from the left.

One possible approach would be to simply estimate the difference in average outcomes between treated and controls within $h$. This strategy is equivalent to fitting a regression including only an intercept at each side of the cutoff. However, since the goal is to estimate two boundary points, this local constant approach will have a bias that can be reduced by including a slope term in the regression. More generally, the most common approach for point estimation  in the continuity-based RD framework is to employ \textit{local polynomial} methods \citep*{Fan-Gijbels_1996_Book}, which involve fitting a polynomial of order $p$ separately on each side of the cutoff, only for observations inside the bandwidth. Local polynomial approximations usually include a weighting scheme that places more weight on observations that are closer to the cutoff; this weighting scheme is based on a \textit{kernel function}, which we denote by $K(\cdot)$. 

More formally, the treatment effect is estimated as:
\[\hat{\tau}=\hat{\alpha}_{+}-\hat{\alpha}_{-}\]
where $\hat{\alpha}_+$ is obtained as the intercept from the (possibly misspecified) regression model:
\[Y_i=\alpha_++\beta_{1+}(X_i-c)+\ldots+\beta_{p+}(X_i-c)^p+u_i\]
on the treated observations using weights $K((X_i-c)/h)$, and similarly $\hat{\alpha}_-$ is obtained as the intercept from an analogous regression fit employing only the control observations. Although theoretically a large value of $p$ can capture more features of the unobserved regression functions, $\mu_1(x)$ and $\mu_0(x)$, in practice high-order polynomials can have erratic behavior, especially when estimating boundary points, a fact usually known as \textit{Runge's phenomenon} \citep*[][p. 1756-57]{Calonico-Cattaneo-Titiunik_2015_JASA}. In addition, global polynomials can lead to counter-intuitive weighting schemes, as discussed by \cite{Gelman-Imbens_2019_JBES}. Common choices for $p$ are $p=1$ or $p=2$.

As we can see, once the bandwidth has been appropriately chosen, the implementation of local polynomial regression reduces to simply fitting two linear or quadratic regressions via weighted least-squares---see \citet*{Cattaneo-Idrobo-Titiunik_2019_Book1} for an extended discussion and practical introduction. Despite the implementation and algebraic similarities between ordinary least squares (OLS) methods and local polynomial methods, there is a crucial difference: OLS methods assume that the polynomial used for estimation is the true form of the function, while local polynomial methods see it as just an approximation to an unknown regression function. Thus, inherent in the use of local polynomial methods is the idea that the resulting estimate will contain a certain error of approximation or misspecification bias. 

This difference between OLS and local polynomial methods turns out to be very consequential for inference purposes---that is, for testing statistical hypotheses and constructing confidence intervals. The conventional OLS inference procedure to test the null hypothesis of no treatment effect at the cutoff, $\mathsf{H_0}: \tau(c)=0$, relies on the assumption that the distribution of the t-statistic is approximately standard normal in large samples:
\begin{equation}
\label{eq:conv} 
\frac{\hat{\tau}}{\sqrt{\mathscr{V}}}\overset{a}{\thicksim} \mathcal{N}(0,1),
\end{equation}
where $\mathscr{V}$ is the (conditional) variance of $\hat{\tau}$, that is, the square of the standard error.

However, this will only occur in cases where the misspecification bias or approximation error of the estimator $\hat{\tau}$ for $\tau(c)$ becomes sufficiently small in large samples, so that the distribution of the t-statistic is correctly centered at zero. In general, this will not occur in RD analysis, where the local polynomials are used as a nonparametric approximation device, and do not make any specific functional form assumptions about the regression functions $\mu_1(x)$ and $\mu_0(x)$, which will be generally misspecified. The general approximation to the t-statistic in the presence of misspecification error is 
\begin{equation}
\label{eq:robust} 
\frac{\hat{\tau} - \mathscr{B}}{\sqrt{\mathscr{V}}}\overset{a}{\thicksim} \mathcal{N}(0,1),
\end{equation}
where $\mathscr{B}$ is the (conditional) bias of $\hat{\tau}$ for $\tau(c)$. This approximation will be equivalent to the one in \eqref{eq:conv} only when $\mathscr{B}/\sqrt{\mathscr{V}}$ is small, at least in large samples.

More generally, it is crucial to account for the bias $\mathscr{B}$ when conducting inference. The magnitude of the bias depends on the shape of the true regression functions and on the length of the bandwidth. As discussed before, the smaller the bandwidth, the smaller the bias. Although the conventional asymptotic approximation in (\ref{eq:conv}) will be valid in some special cases, such as when the bandwidth is small enough, it is not valid in general. In particular, if the bandwidth chosen for implementation is the MSE-optimal bandwidth discussed in the prior section, the bias will remain even in large samples, making inferences based on (\ref{eq:conv}) invalid. In other words, the MSE-optimal bandwidth, which is optimal for point estimation, is too large when conducting inference according to the usual OLS approximations. 

Generally valid inferences thus require researchers to use the asymptotic approximation in (\ref{eq:robust}), which contains the bias. In particular, \citet*{Calonico-Cattaneo-Titiunik_2014_ECMA} propose a way to construct a t-statistic that corrects the bias of the estimator (thus making the approximation valid for more bandwidth choices, including the MSE-optimal choice) and simultaneously adjusts the standard errors to account for the variability that is introduced in the bias correction step---this additional variability is introduced because the bias is unknown and thus must be estimated. This approach is known as \textit{robust bias-corrected} inference.

Based on the approximation (\ref{eq:robust}), \citet*{Calonico-Cattaneo-Titiunik_2014_ECMA}  propose robust bias-corrected confidence intervals
\[\mathtt{CI}_{\mathtt{rbc}}
= \left[ ~ \big(\hat{\tau}-\hat{\mathscr{B}}\big) \pm 1.96 \cdot \sqrt{\mathscr{V}_{\mathtt{bc}}} ~ \right], \]
where, in general,  $\mathscr{V}_{\mathtt{bc}} > \mathscr{V}$ because $\mathscr{V}_{\mathtt{bc}}$ includes the variability of estimating $\mathscr{B}$ with $\hat{\mathscr{B}}$. In terms of implementation, the infeasible variance $\mathscr{V}_{\mathtt{bc}}$ can be replaced by a consistent estimator $\hat{\mathscr{V}}_{\mathtt{bc}}$, which can account for heteroskedasticity and clustering as apprpriate.

Robust bias correction methods for RD designs have been further developed in recent years. For example, see \citet*{Calonico-Cattaneo-Farrell-Titiunik_2019_RESTAT} for robust bias correction inference in the context of RD designs with covariate adjustments, clustered data, and other empirically relevant features. In addition, see \citet*{Calonico-Cattaneo-Farrell_2018_JASA,Calonico-Cattaneo-Farrell_2019_CEoptimal,Calonico-Cattaneo-Farrell_2019_CERD} for theoretical results justifying some of features of robust bias correction inference. Finally, see \citet*{Ganong-Jager_2018_JASA} and \citet*{Hyytinen-etal_2018_QE} for two recent applications and empirical comparisons of robust bias correction methods.\bigskip

\begin{tcolorbox}[colframe=blue!25,
colback=blue!10,
coltitle=blue!20!black,
title = \textbf{Continuity-based framework: summary}]

\begin{enumerate}
\item \textbf{Key assumptions}:
	\begin{enumerate}
	\item Random potential outcomes drawn from an infinite population
	\item The regression functions are continuous at the cutoff
	\end{enumerate}
\item \textbf{Bandwidth selection}:
	\begin{enumerate}
	\item Systematic, data-driven selection based on non-parametric methods
	\item Optimality criteria: MSE, coverage error
	\end{enumerate}
\item \textbf{Estimation}:
	\begin{enumerate}
	\item Nonparametric local polynomial regression within bandwidth
	\item Choice parameters: order of the polynomial, weighting method (kernel)
	\end{enumerate}
\item \textbf{Inference}:
	\begin{enumerate}
	\item Large-sample normal approximation
	\item Robust, bias corrected
	\end{enumerate}
\end{enumerate}
\end{tcolorbox}


\section{The Local Randomization Framework} 

The local randomization approach to RD analysis provides an alternative to the continuity-based framework. Instead of relying on assumptions about the continuity of regression functions and their approximation and extrapolation, this approach is based on the idea that, close enough to the cutoff, the treatment can be interpreted to be ``as good as randomly assigned''. The intuition is that, if units either have no knowledge of the cutoff or have no ability to precisely manipulate their own score, units whose scores are close enough to the cutoff will have the same chance of being barely above the cutoff as barely below it. If this is true, close enough to the cutoff, the RD design may create experimental-like variation in treatment assignment. The idea that RD designs create conditions that resemble an experiment near the cutoff has been present since the origins of the method \citep[see][]{Thistlethwaite-Campbell_1960_JEP}, and has been sometimes proposed as a heuristic interpretation of continuity-based RD results.

\citet*{Cattaneo-Frandsen-Titiunik_2015_JCI} used this local randomization idea to develop a formal framework, and to derive alternative assumptions for the analysis of RD designs, which are stronger than the typical continuity conditions. The formal local randomization framework was further developed by \citet*{Cattaneo-Titiunik-VazquezBare_2017_JPAM}. The central idea behind the local randomization approach is to assume the existence of a neighborhood or window around the cutoff where the assignment to being above or below the cutoff behaves as it would have behaved in an actual experiment. In other words, the local randomization RD approach makes the assumption that there is a window around the cutoff where assignment to treatment is as-if experimental. 
 
The formalization of these assumptions requires a more general notation. In prior sections, we used $Y_i(D_i)$ to denote the potential outcome under treatment $D_i$, which could be equal to one (treatment) or zero (control). Since $D_i = \I(X_i \geq c)$, this also allowed the score $X_i$ to indirectly affect the potential outcomes; moreover, this notation did not prevent $Y_i(\cdot)$ from being a function of $X_i$, but this was not explicitly noted. We now generalize the notation to explicitly note that the potential outcomes may be a direct function of $X_i$, so we write  $Y_i(D_i,X_i)$. In addition, note that here and in all prior sections we are implicitly assuming that potential outcomes only depend on unit $i$'s own treatment assignment and running variable, an assumption known as SUTVA (stable unit treatment value assumption). While some of the methods described in this section are robust to some violations of the SUTVA, we impose this assumption to ease exposition. See \citet*{Cattaneo-Titiunik-VazquezBare_2017_JPAM} for more discussion.

To formalize the local randomization RD approach, we assume that there exists a window $W_0$ around the cutoff where the following two conditions hold:

 \begin{itemize}
 	\item \textbf{Unconfounded Assignment}. The distribution function of the score inside the window, $F_{X_i | X_i \in W_0}(r)$, does not depend on the potential outcomes, is the same for all units, and is known: 
 	\begin{equation}
 	\label{eq:uncon}
 	F_{X_i | X_i \in W_0}(x) = F_0(x),
 	\end{equation}
 	where $F_0(x)$ is a known distribution function.
 	
 	\item \textbf{Exclusion Restriction}. The potential outcomes do not depend on the value of the running variable inside the window, except via the treatment assignment indicator
 	\begin{equation}\label{eq:flatout}
 	Y_i(d,x)=Y_i(d)\quad \forall\, i \text{ such that } X_i \in W_0\text{.}
 	\end{equation}
 	This condition requires the potential outcomes to be unrelated to the score inside the window. 
 \end{itemize}	
 
Importantly, these two assumptions would not be satisfied by randomly assigning the treatment inside $W_0$, because the random assignment of $D_i$ inside $W_0$ does not by itself guarantee that the score and the potential outcomes are unrelated (the exclusion restriction). For example, imagine a RD design based on elections, where the treatment is the electoral victory of a political party, the score is the vote share, and the party wins the election if the vote share is above 50\%. Even if, in very close races, election winners were chosen randomly instead of based on their actual vote share, donors might still believe that districts where the party obtained a bare majority are more likely to support the party again, and thus they may donate more money to the races where the party's vote share was just above 50\% than to races where the party was just below 50\%. If donations are effective in boosting the party, this would induce a positive relationship near the cutoff between the running variable (vote share) and the outcome of interest (victory in the future election).

The discussion above illustrates why the unconfounded assignment assumption in equation (\ref{eq:uncon}) is not enough for a local randomization approach to RD analysis. We must explicitly assume that the score and the potential outcomes are unrelated inside $W_0$, which is not implied by (\ref{eq:uncon}). This issue is discussed in detail by \citet*{Sekhon-Titiunik_2017_AIE}, who use several examples to show that the exclusion restriction in (\ref{eq:flatout}) is neither implied by assuming statistical independence between the potential outcomes and the treatment in $W_0$, nor by assuming that the running variable is randomly assigned in $W_0$. In addition, see \citet*{Sekhon-Titiunik_2016_ObsStud} for a discussion of the status of RD designs among observational studies, and \cite{Titiunik2020-Chapter} for a discussion of the connection between RD designs and natural experiments.

\subsection{Estimation and Inference within a Known Window}\label{sec:estinf2}

The local randomization conditions (\ref{eq:uncon}) and (\ref{eq:flatout}) open new possibilities for RD estimation and inference. Of course, these conditions are strong and, just like the continuity conditions in Section \ref{sec:continuity}, they are not implied by the RD treatment assignment rule but rather must be assumed in addition to it \citep{Sekhon-Titiunik_2016_ObsStud}. Because these assumptions are strong and are inherently untestable, it is crucial for researchers to provide as much information as possible regarding their plausibility. We discuss this issue in Section \ref{sec:falsification}, where we present several strategies for empirical falsification of the RD assumptions. 

The key assumption of the local randomization approach is that there exists a neighborhood around the cutoff in which (\ref{eq:uncon}) and (\ref{eq:flatout}) hold---implying that we can treat the RD design as a randomized experiment near the cutoff. We denote this neighborhood by $W_0=[c-w,c+w]$, where $c$ continues to be the RD cutoff, but we now use the notation $w$ as opposed to $h$ to emphasize that $w$ will be chosen and interpreted differently from the previous section. Furthermore, to ease the exposition, we start by assuming that $W_0$ is known, and then discuss how to select $W_0$ based on observable information. This data-driven window selection step will be crucial in applications, as in most empirical examples $W_0$ is fundamentally unknown, if it exists at all---but see \cite{Hyytinen-etal_2018_QE} for an exception.

Given a window $W_0$, the local randomization framework summarized by assumptions (\ref{eq:uncon}) and (\ref{eq:flatout}) allows us to analyze the RD design employing the standard tools of the classical analysis of experiments. Depending on the available number of observations inside the window, the experimental analysis can follow two different approaches. In the Fisherian approach, also known as a randomization inference approach, potential outcomes are considered non-random, the assignment mechanism is assumed to be known, and this assignment is used to calculate the exact finite-sample distribution of a test statistic of interest under the null hypothesis that the treatment effect is zero for every unit. On the other hand, in the large-sample approach, the potential outcomes may be fixed or random, the assignment mechanism need not be known, and the finite-sample distribution of the test statistic is approximated under the assumption that the number of observations is large. Thus, in contrast to the Fisherian approach, in the large-sample approach inferences are based on test statistics whose finite-sample properties are unknown, but whose null distribution  can be approximated by a Normal distribution under the assumption that the sample size is large enough.

Below we briefly review both Fisherian and large-sample methods for analysis of RD designs under a local randomization framework. Fisherian methods will be most useful when the number of observations near the cutoff is small, which may render large-sample methods invalid. In contrast, in applications with many observations, large-sample methods will be the most natural approach, and Fisherian methods can be used as a robustness check.

\subsubsection{Fisherian approach}
In the Fisherian framework, the potential outcomes are seen as fixed, non-random magnitudes from a finite population of $n$ units. The information on the observed sample of units $i=1,\ldots,n$ is not seen as a random draw from an infinite population, but as the population of interest. This feature allows for the derivation of the finite-sample-exact distribution of test statistics without relying on approximations.

We follow the notation in \citet*{Cattaneo-Titiunik-VazquezBare_2017_JPAM}, adapting slightly our previous notation. Let $\mathbf{X}=(X_1,\ldots,X_n)'$ denote the $n\times 1$ column vector collecting the observed running variable of all units in the sample, and $\mathbf{D}=(D_1,\ldots,D_n)'$ be the vector collecting treatment assignments. The non-random potential outcomes for each unit $i$ are denoted by $y_i(d,x)$ where $d$ and $x$ are possible values for $D_i$ and $X_i$. All the potential outcomes are collected in the vector $\mathbf{y}(\mathbf{d},\mathbf{x})$. The vector of observed outcomes is simply the vector of potential outcomes, evaluated at the observed values of the treatment and running variable, $\mathbf{Y}=\mathbf{y}(\mathbf{D},\mathbf{X})$.

Because potential outcomes are assumed non-random, all the randomness in the model enters through the running variable vector $\mathbf{X}$, and the treatment assignment $\mathbf{D}$ which is a function of it. In what follows, we let the subscript ``0'' indicate the subvector inside the neighborhood $W_0$, so that $\mathbf{X}_0$, $\mathbf{D}_0$ and $\mathbf{Y}_0$ denote the vectors of running variables, treatment assignments and observed outcomes inside $W_0$. Finally, $N_0^+$ will denote the number of observations inside the neighborhood and above the cutoff (treated units inside $W_0$), and $N_0^-$ the number of units in the neighborhood below the cutoff (control units in $W_0$), with $N_0=N_0^++N_0^-$.  Note that using the fixed-potential outcomes notation, the exclusion restriction becomes  $y_i(d,x)=y_i(d),\; \forall\, i \text{ in } W_0$ \citep*[see assumption 1(b) in][]{Cattaneo-Frandsen-Titiunik_2015_JCI}.  

In this Fisherian framework, a natural null hypothesis to test for the presence of a treatment effect is the \textit{sharp null of no effect}:
\[\mathsf{H}^s_0:\quad y_i(1)=y_i(0), \quad \forall i \text{ in } W_0\text{.}\]
This sharp null hypothesis states that switching treatment status does not affect potential outcomes, implying that the treatment does not have an effect on \textit{any} unit inside the window. In this context, a hypothesis is \textit{sharp} when it allows the researcher to impute all the missing potential outcomes. Thus, $\mathsf{H}^s_0$ is sharp because when there is no effect, all the missing potential outcomes are equal to the observed ones. 

Under $\mathsf{H}^s_0$, the researcher can impute all the missing potential outcomes and, since the assignment mechanism is assumed to be known, it is possible to calculate the distribution of any test statistic $T(\mathbf{D}_0,\mathbf{Y}_0)$ to assess how far in the tails the observed statistic falls. This reasoning provides a way to calculate a p-value for $\mathsf{H}^s_0$ that is finite-sample exact and does not require any distributional approximation. This randomization inference p-value is obtained by calculating the value of $T(\mathbf{D}_0,\mathbf{Y}_0)$ for all possible values of the treatment vector inside the window $\mathbf{D}_0$, and calculating the probability of $T(\mathbf{D}_0,\mathbf{Y}_0)$ being larger than the observed value $T_\mathsf{obs}$. See \citet*{Cattaneo-Frandsen-Titiunik_2015_JCI}, \citet*{Cattaneo-Titiunik-VazquezBare_2017_JPAM} and \citet*{Cattaneo-Titiunik-VazquezBare_2016_Stata} for further details and implementation issues. See also \citet*{Cattaneo-Idrobo-Titiunik_2019_Book2} for a practical introduction to local randomization methods.

In addition to testing the null hypothesis of no treatment effect, the researcher may be interested in obtaining a point estimate for the effect. When condition (\ref{eq:flatout}) holds, a difference in means between treated and controls inside the window,
\[\Delta=\frac{1}{N^+_0}\sum_{i=1}^n Y_iD_i - \frac{1}{N^-_0}\sum_{i=1}^n Y_i(1-D_i),\]
where the sum runs over all observations inside $W_0$, is unbiased for the sample average treatment effect in $W_0$,
\[\tau_0=\frac{1}{N_0}\sum_{i=1}^n (y_i(1)-y_i(0)).\]

However, it is important to emphasize that the randomization inference method described above cannot test hypotheses on $\tau_0$ because the null hypothesis that $\tau_0=0$ is not sharp, that is, does not allow the researcher to unequivocally impute all the missing potential outcomes, without further restrictive assumptions, which is a necessary condition to use Fisherian methods. Hence, under the assumptions imposed so far, hypothesis testing on $\tau_0$ has to be based on asymptotic approximations, as described in Section \ref{sec:LR-largesample}.

The assumption that the potential outcomes do not depend on the running variable, stated in Equation (\ref{eq:flatout}), can be relaxed by assuming a local parametric model for the relationship between $\mathbf{Y}_0$ and $\mathbf{X}_0$. Specifically, \citet*{Cattaneo-Titiunik-VazquezBare_2017_JPAM} assume there exists a transformation $\phi(\cdot)$ such that the transformed outcomes do not depend on $\mathbf{X}_0$. This transformation could be, for instance, a linear adjustment that removes the slope whenever the relationship between outcomes and the running variable is assumed to be linear. The case where potential outcomes do not depend on the running variable is a particular case in which $\phi(\cdot)$ is the identity function. Both inference and estimation can therefore be conducted using the transformed outcomes when the assumption that potential outcomes are unrelated is not reasonable, or as a robustness check.

\subsubsection{Large-Sample approach}
\label{sec:LR-largesample}
In the most common large-sample approach, we treat potential outcomes as random variables, and often see the units in the study as a random sample from a larger population. (Though in the Neyman large-sample approach, potential outcomes are fixed; see \cite{Imbens-Rubin_2015_Book} for more discussion.) In addition to the randomness of the potential outcomes, this approach differs from the Fisherian approach in its null hypothesis of interest. Given the randomness of the potential outcomes, the focus is no longer on the sharp null but rather typically on the hypothesis that the average treatment effect is zero. In our RD context, this null hypothesis can be written as
\[\mathsf{H}^s_0:\quad \E[Y_i(1)]=\E[Y_i(0)], \quad \forall i \text{ in } W_0\]

Inference in this case is based on the usual large-sample methods for the analysis of experiments, relying on usual difference-in-means tests and Normal-based confidence intervals. See \cite{Imbens-Rubin_2015_Book} and \cite{Cattaneo-Idrobo-Titiunik_2019_Book2} for details.

\subsection{Window Selection}

In practice, the window $W_0$ in which the RD design can be seen as a randomized experiment is not known and needs to be estimated. \citet*{Cattaneo-Frandsen-Titiunik_2015_JCI} propose a window selection mechanism based on the idea that in a randomized experiment, the distribution of observed covariates has to be equal between treated and controls. Thus, if the local assumption is plausible in any window, it should be in a window where we cannot reject that the pre-determined characteristics of treated and control units are on average identical. 

The idea of this procedure is to select a test statistic that summarizes differences in a vector of covariates between groups, such as difference-in-means or the Kolmogorov-Smirnov statistic, and start with an initial ``small'' window. Inside this initial window, the researcher conducts a test of the null hypothesis that covariates are balanced between treated and control groups. This can be done, for example, by assessing whether the minimum p-value from the tests of differences-in-means for each covariate is larger than some specified level, or by conducting a joint test using for instance a Hotelling statistic. If the null hypothesis is not rejected, enlarge the window and repeat the process. The selected window will be the widest window in which the null hypothesis is not rejected. Common choices for the test statistic $T(\mathbf{D}_0,\mathbf{Y}_0)$ are the difference-in-means between treated and controls, the two-sample Kolmogorov-Smirnov statistic or the rank sum statistic. The minimum window to start the procedure should contain enough observations to ensure enough statistical power to reject the null hypothesis of covariate balance. The appropriate minimum number of observations will naturally depend on unknown, application-specific parameters, but based on standard power calculations we suggest using no less than approximately 10 observations in each group.

See \citet*{Cattaneo-Frandsen-Titiunik_2015_JCI} and \citet*{Cattaneo-Titiunik-VazquezBare_2017_JPAM} for methodological details, \citet*{Cattaneo-Idrobo-Titiunik_2019_Book2} for a practical introduction, and \citet*{Cattaneo-Titiunik-VazquezBare_2016_Stata} for software implementation.

\bigskip
\begin{tcolorbox}[colframe=blue!25,
colback=blue!10,
coltitle=blue!20!black,
title=\textbf{Local randomization framework: summary}]

\begin{enumerate}
\item \textbf{Key assumptions}:
	\begin{enumerate}
	\item There exists a window $W_0$ in which the treatment assignment mechanism satisfies two conditions:
	\begin{itemize}
    \item Probability of receiving a particular score value in 	$W_0$ does not depend on the potential outcomes
	\item Exclusion restriction or parametric relationship between $\mathbf{Y}$ and $\mathbf{X}$ in $W_0$
	\end{itemize}
	\end{enumerate}
\item \textbf{Window selection}:
	\begin{enumerate}
	\item Goal: Find a window where the key assumptions are plausible
	\item Iterative procedure to balance observed covariates between groups
	\item Choice parameters: test statistic, stopping rule
	\end{enumerate}
\item \textbf{Estimation}:
	\begin{enumerate}
	\item Difference in means between treated and controls within neighborhood OR
	\item Flexible parametric modeling to account for the effect of $X_i$
	\end{enumerate}
\item \textbf{Inference}:
	\begin{enumerate}
	\item Fisherian randomization-based inference or large-sample inference
	\item Conditional on sample and chosen window
	\item Choice parameter: test statistic, randomization mechanism in Fisherian
	\end{enumerate}
\end{enumerate}
\end{tcolorbox}

\section{Falsification Methods}\label{sec:falsification}

Every time researchers use an RD design, they must rely on identification assumptions that are fundamentally untestable, and that do not hold by construction. If we employ a continuity-based approach, we must assume that the regression functions are smooth functions of the score at the cutoff. If, on the other hand, we employ a local randomization approach, we must assume that there exists a window where the treatment behaves as if it had been randomly assigned.  These assumptions may be violated for many reasons. Thus, it is crucial for researchers to provide as much empirical evidence as possible about its validity. 

Although testing the assumptions directly is not possible, there are several empirical regularities that we expect to hold in most cases where the assumptions are met. We discuss some of these tests below. Our discussion is brief, but we refer the reader to \citet*{Cattaneo-Idrobo-Titiunik_2019_Book1} for an extensive practical discussion of RD falsification methods, and additional references. 

\begin{enumerate}
	\item \textbf{Covariate Balance}. If either the continuity or local randomization assumptions hold, the treatment should not have an effect on any predetermined covariates, that is, on covariates whose values are realized before the treatment is assigned. Since the treatment effect on predetermined  covariates is zero by construction, consistent evidence of non-zero effects on covariates that are likely to be confounders would raise questions about the validity  of the RD assumptions. For implementation, researchers should analyze each covariate as if it were an outcome. In the continuity-based approach, this requires choosing a bandwidth and performing local polynomial estimation and inference within that bandwidth. Note that the optimal bandwidth is naturally different for each covariate. In the local randomization approach, the null hypothesis of no effect should be tested for each covariate using the same choices as used for the outcome. If the window is chosen using the covariate balance procedure discussed above, the selected window will automatically be a region where no treatment effects on covariates are found. 
	
    \item \textbf{Density of Running Variable}. Another common falsification test is to study the number of observations near the cutoff. If units cannot manipulate precisely the value of the score that they receive, we should expect as many observations just above the cutoff as just below it. In contrast, for example, if units had the power to affect their score and they knew that the treatment were very beneficial, we should expect more people just above the cutoff (where the treatment is received) than below it. In the continuity-based framework, the procedure is to test the null hypothesis that the density of the running variable is continuous at the cutoff \citep{McCrary_2008_JoE}, which can be implemented in a more robust way via the novel density estimator proposed in \citet*{Cattaneo-Jansson-Ma_2019_JASA}. In the local randomization framework, \citet*{Cattaneo-Titiunik-VazquezBare_2017_JPAM} propose a novel implementation via a finite sample exact binomial test of the null hypothesis that the number of treated and control observations in the chosen window is compatible with a 50\% probability of treatment assignment.
    
    \item \textbf{Alternative cutoff values}. Another falsification test estimates the treatment effect on the outcome at a cutoff value different from the actual cutoff used for the RD treatment assignment, using the same procedures used to estimate the effect in the actual cutoff but only using observations that share the same treatment status (all treatment observations if the artificial cutoff is above the real one, or all control observations if the artificial cutoff is below the real cutoff). The idea is that no treatment effect should be found at the artificial cutoff, since the treatment status is not changing. 
    
  	\item \textbf{Alternative bandwidth and window choices}. Another approach is to study the robustness of the results to small changes in the size of the bandwidth or window. For implementation, the main analysis is typically repeated for values of the bandwidth or window that are slightly smaller and/or larger than the values used in the main analysis. If the effects completely change or disappear for small changes in the chosen neighborhood, researchers should be cautious in interpreting their results.
        
\end{enumerate}

\section{Empirical Illustration}

To illustrate all the RD methods discussed so far, we partially re-analyze the study by \citet*{KlasnjaTitiunik2017-APSR}. These authors study municipal mayor elections in Brazil between 1996 and 2012, examining the effect of a party's victory in the current election on the probability that the party wins a future election for mayor in the same municipality. The unit of analysis is the municipality, the score is the party's margin of victory at election $t$---defined as the party's vote share minus the vote share of the party's strongest opponent, and the treatment is the party's victory at $t$. Their original analysis focuses on the unconditional victory of the party at $t+1$ as the outcome of interest. In this illustration, our outcome of interest is instead the party's margin of victory at $t+1$, which is only defined for those municipalities where the incumbent party runs for reelection at $t+1$. We analyze this effect for the incumbent party (defined as the that party won election $t-1$, whatever this party is) in the full sample. \citet*{KlasnjaTitiunik2017-APSR} discuss the interpretation and validity issues that arise when conditioning on the party's decision to re-run, but we ignore such issues here for the purposes of illustration.

In addition to the outcome and score variables used for the main empirical analysis, our covariate-adjusted local polynomial methods, window selection procedure, and falsification approaches employ seven covariates at the municipality level: per-capita GDP, population, number of effective parties, and indicators for whether each of four parties (the Democratas, PSDB, PT and PMDB) won the prior ($t-1$) election. 

We implement the continuity-based analysis with the \texttt{rdrobust} software \citep*{Calonico-Cattaneo-Titiunik_2014_Stata,Calonico-Cattaneo-Titiunik_2015_R,Calonico-Cattaneo-Farrell-Titiunik_2017_Stata}, the local randomization analysis using the \texttt{rdlocand} software \citep*{Cattaneo-Titiunik-VazquezBare_2016_Stata}, and the density test falsification using the \texttt{rddensity} software \citep*{Cattaneo-Jansson-Ma_2018_Stata}. The packages can be obtained for \texttt{R} and \texttt{Stata} from \url{https://sites.google.com/site/rdpackages/}. We do not present the code to conserve space, but the full code employed is available in the packages' website. \citet*{Cattaneo-Idrobo-Titiunik_2019_Book1,Cattaneo-Idrobo-Titiunik_2019_Book2} offer a detailed tutorial on how to use these packages, employing a different empirical illustration.

\subsection*{Falsification Analysis}

\begin{figure}[H]
	\centering
	\caption{Estimated density of running variable}
	\includegraphics[width=0.50\textwidth]{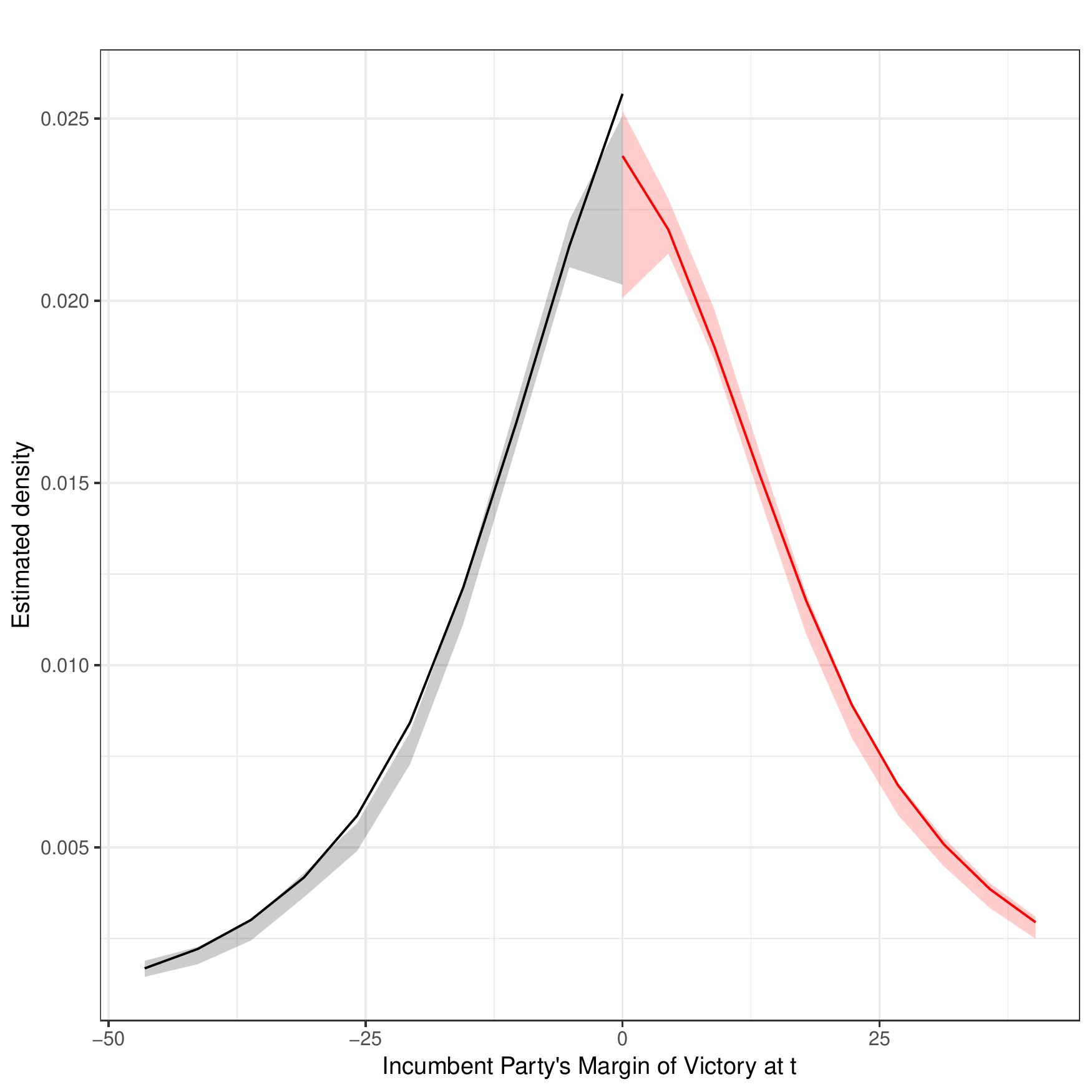}
	\label{fig:rddensity}
\end{figure}

We start by presenting a falsification analysis. In order to falsify the continuity-based analysis, we analyze the density of the running variable, and also the effect of the RD treatment on several predetermined covariates.  We start by reporting the result of a continuity-based density test, using the local polynomial density estimator developed by \citet*{Cattaneo-Jansson-Ma_2019_JASA}. The estimated difference in the density of the running variable at the cutoff is $-0.0753$, and the p-value associated with the test of the null hypothesis that this difference is zero is $0.94$. This test is illustrated in Figure \ref{fig:rddensity}, which shows the local-polynomial-estimated density of the incumbent party's margin of victory at $t$ at the cutoff, separately estimated from above and below the cutoff. These results indicate that the density of the running variable does not change abruptly at the cutoff, and are thus consistent with the assumption that parties do not precisely manipulate their margin of victory to ensure a win in close races.

In addition, we also implemented the finite sample exact binomial tests proposed in \citet*{Cattaneo-Titiunik-VazquezBare_2017_JPAM}, which confirmed the empirical results obtained via local polynomial density methods. We do not report these numerical result to conserve space, but they can be consulted using the accompaying replication files.

We also present local polynomial point estimates of the effect of the incumbent party's victory on each of the seven predetermined covariates mentioned above, and we perform robust local-polynomial inference to obtain confidence intervals and p-values for these effects. Since these covariates are all determined before the outcome of the election at $t$ is known, the treatment effect on each of them is zero by construction. Our estimated effects and statistical inferences should therefore be consistent with these known null effects. 

\begin{figure}[H]
	\centering
	\subfloat[][GDP per capita]{
		\includegraphics[width=0.55\textwidth]{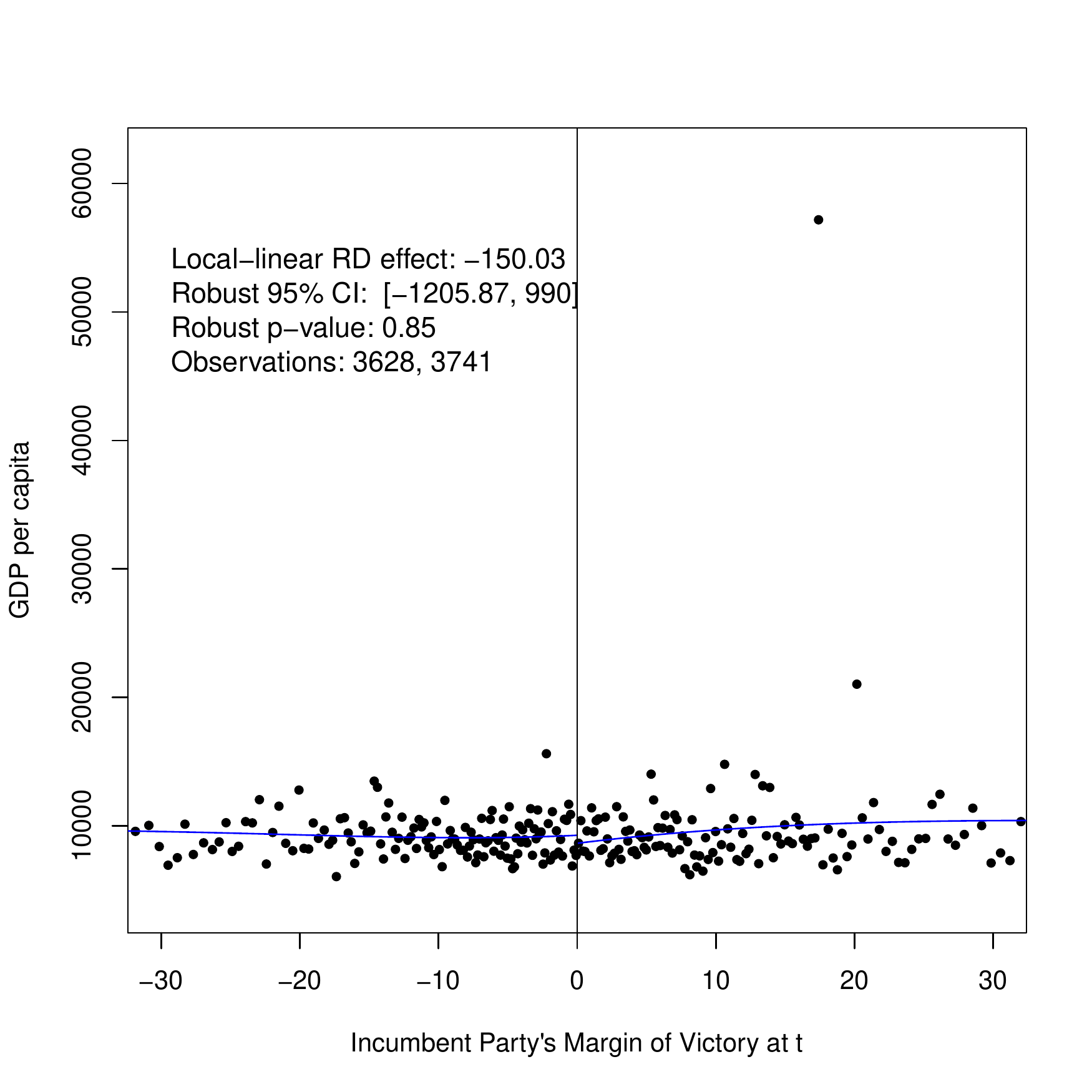}
		\label{fig:covs2GDP}}
	\subfloat[][Population]{
		\includegraphics[width=0.55\textwidth]{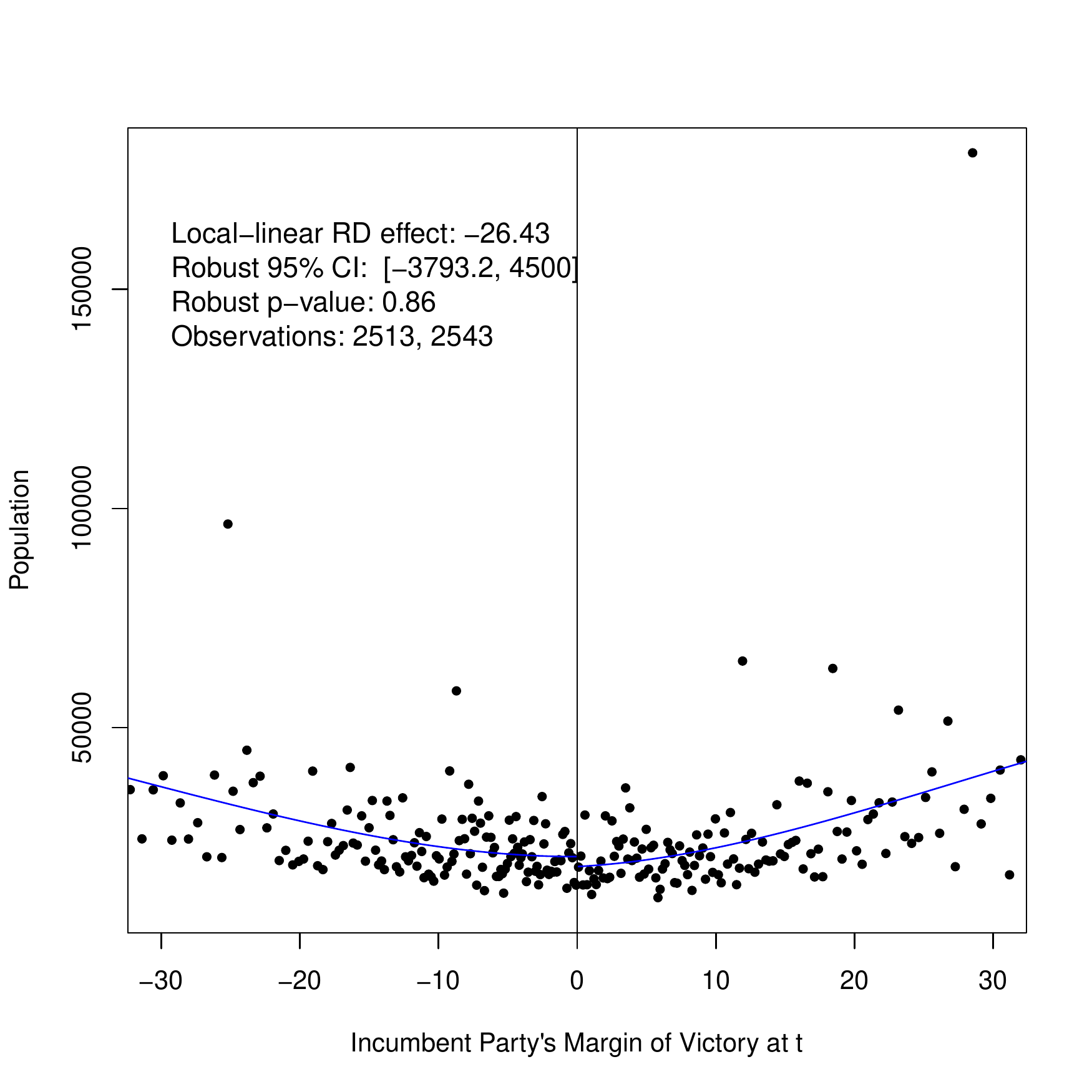}
		\label{fig:covs2Pop}}
	\qquad
	\subfloat[][No. Effective Parties]{
		\includegraphics[width=0.55\textwidth]{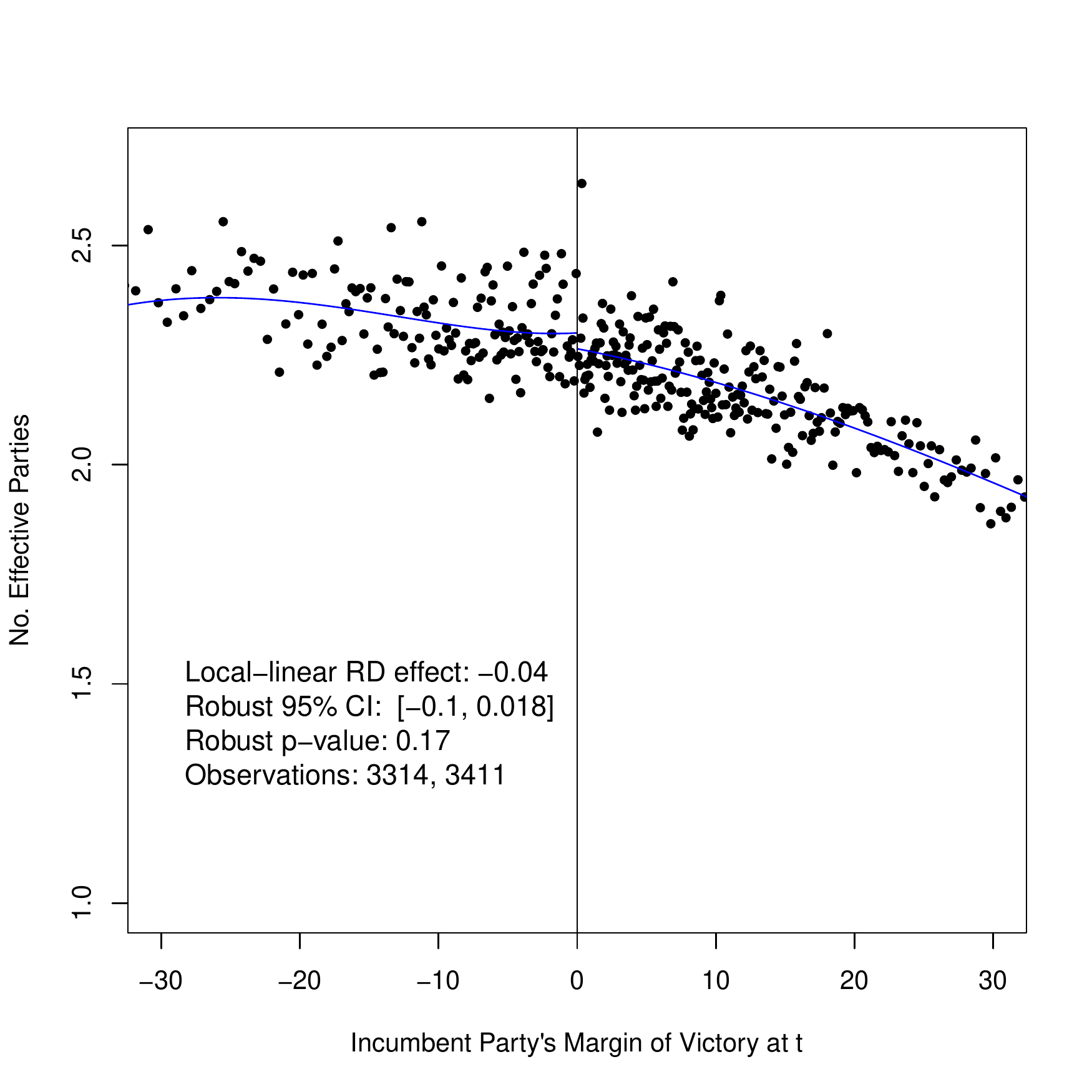}
		\label{fig:covs2EffP}}
	\subfloat[][PMDB Victory at $t-1$]{
		\includegraphics[width=0.55\textwidth]{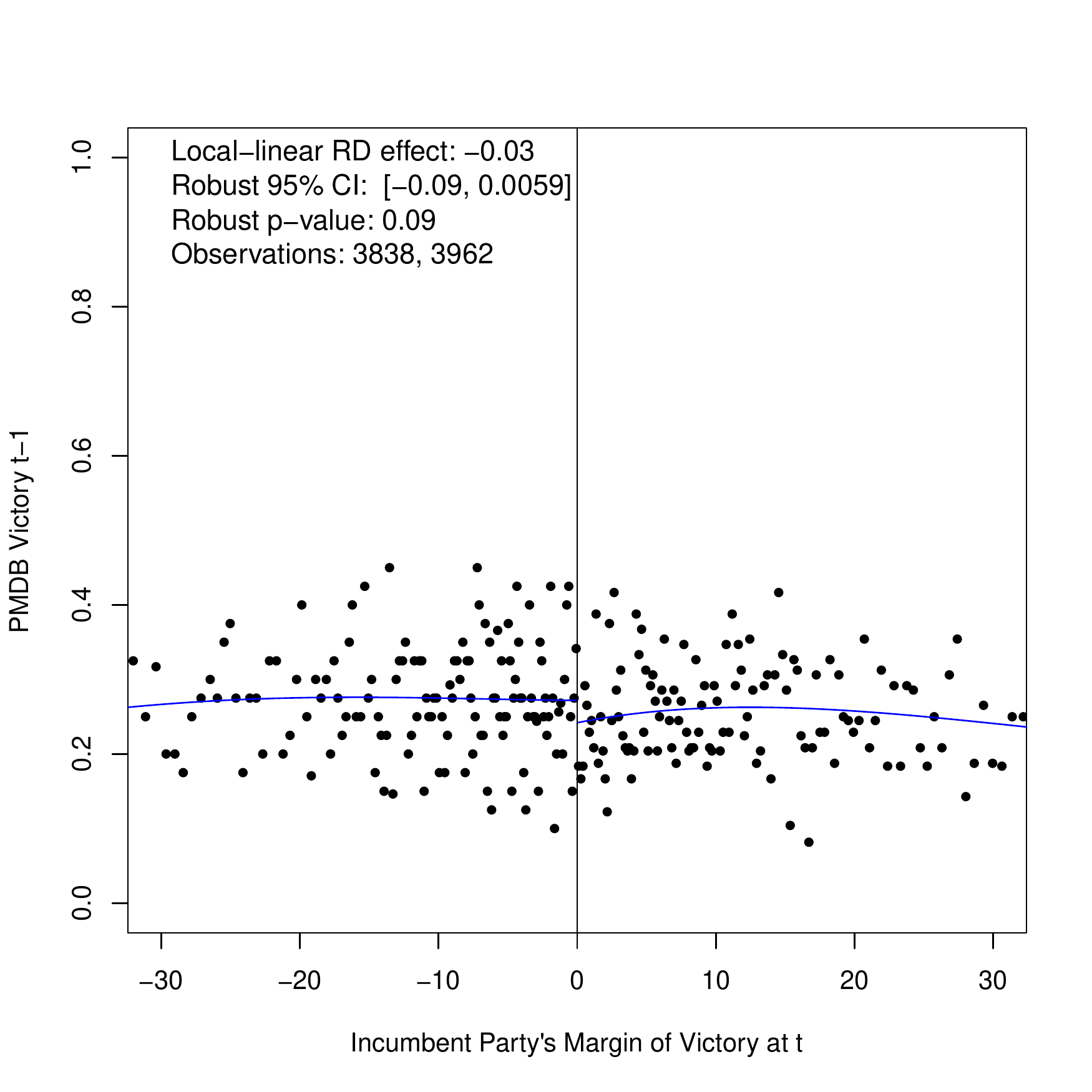}
		\label{fig:covs2PMDB}}
	\caption{RD Effects on Predetermined Covariates}
	\label{fig:covs1}
\end{figure}

\begin{figure}[H]
	\centering
	\subfloat[][DEM Victory at $t-1$]{
		\includegraphics[width=0.55\textwidth]{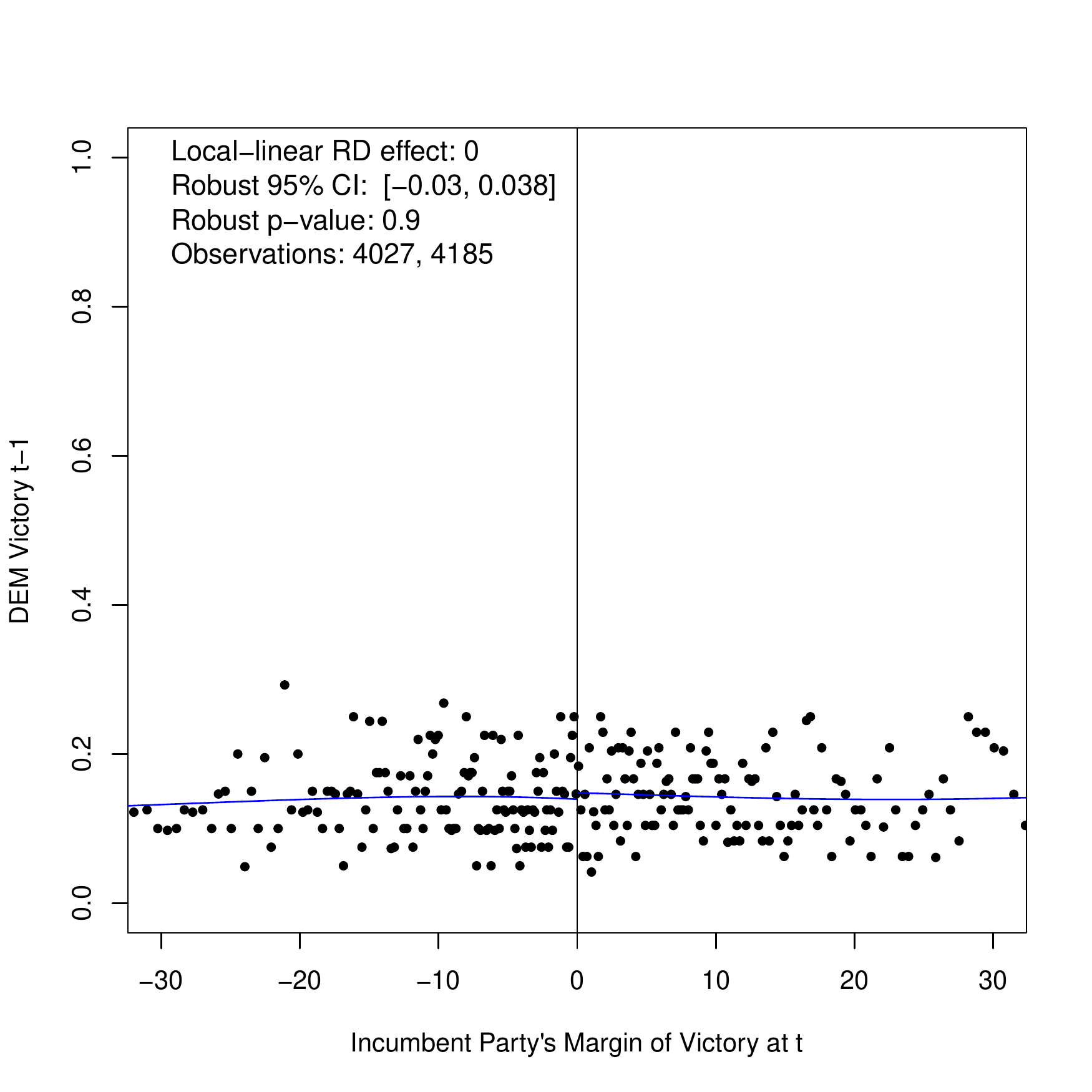}
		\label{fig:covs1DEM}}
	\subfloat[][PSDB Victory at $t-1$]{
		\includegraphics[width=0.55\textwidth]{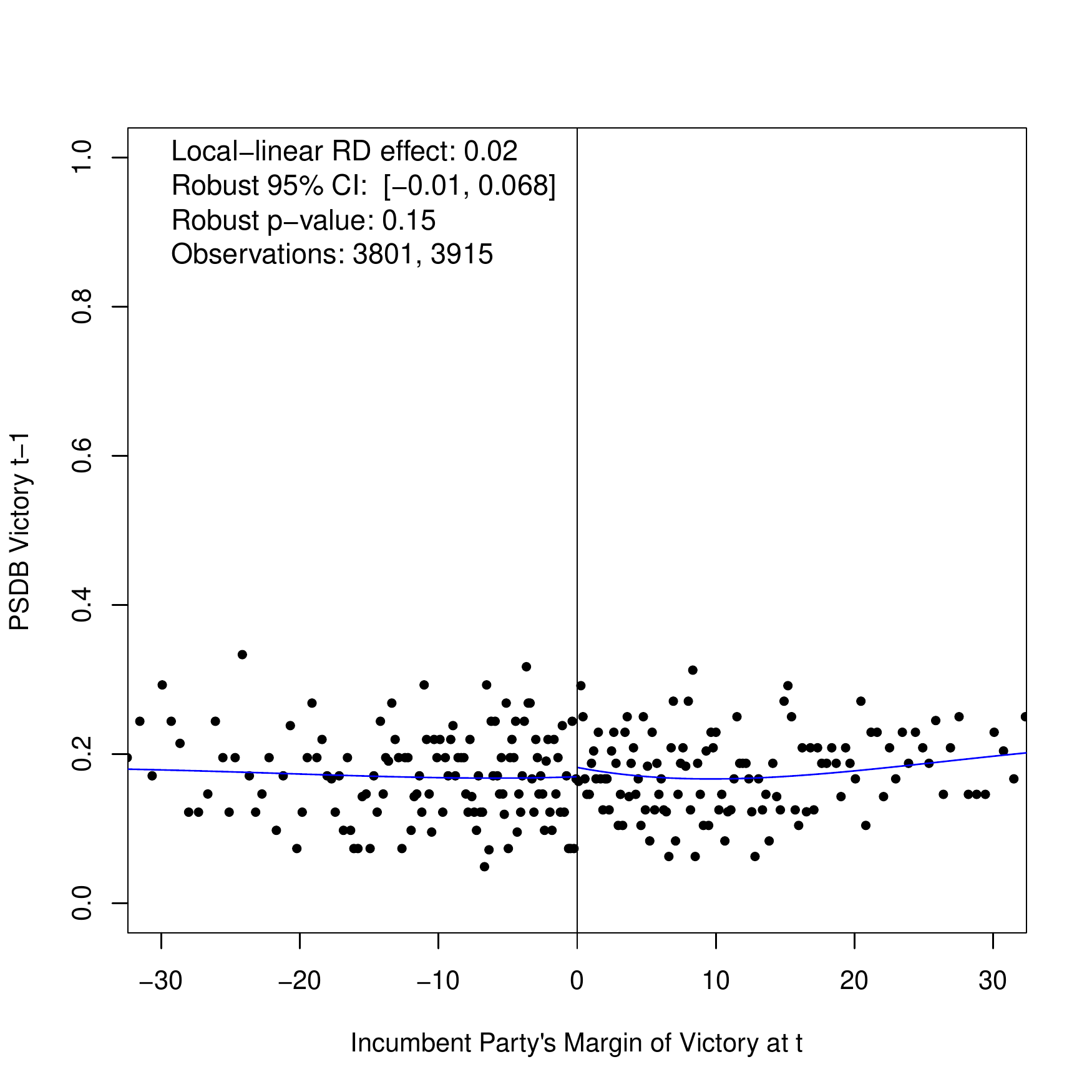}
		\label{fig:covs1PSDB}}
			\qquad
	\subfloat[][PT Victory at $t-1$]{
		\includegraphics[width=0.55\textwidth]{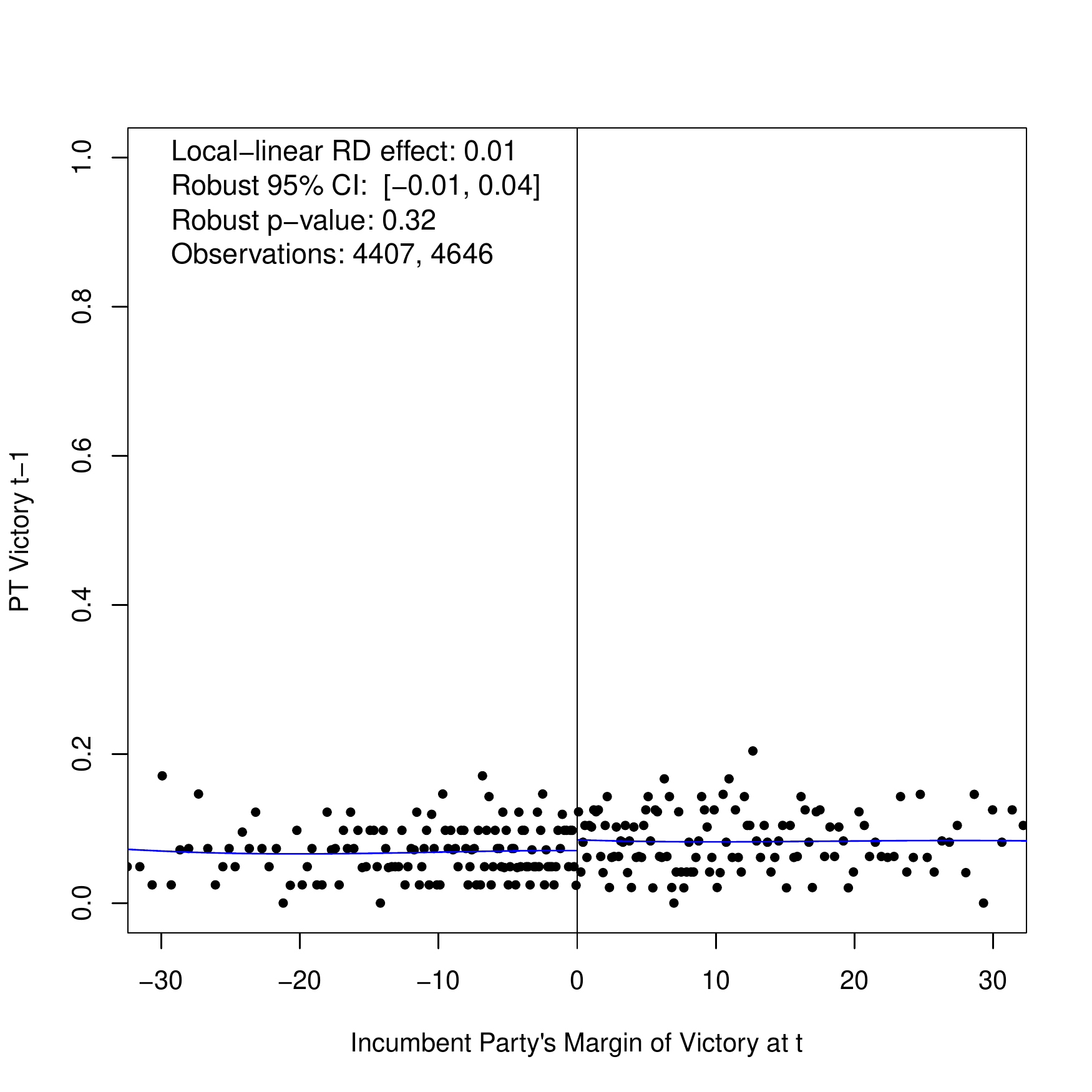}
		\label{fig:covs1PT}}
	\caption{RD Effects on Predetermined Covariates}
	\label{fig:covs2}
\end{figure}

We present the results graphically in Figures \ref{fig:covs1} and \ref{fig:covs2} using typical RD plots \citep*{Calonico-Cattaneo-Titiunik_2015_JASA} where binned means of the outcome within intervals of the score are plotted against the mid point of the score in each interval. A fourth-order polynomial, separately estimated above and below the cutoff, is superimposed to show the global shape of the regression functions. In these plots,  we also report the formal local polynomial point estimate, $95\%$ robust confidence interval, robust p-value, and number of observations within the bandwidth. The bandwidth (not reported) is chosen in each case to be MSE-optimal. 

As we can see, the incumbent party's bare victory at $t$ does not have an effect on any of the covariates. All 95\% confidence intervals contain zero, most of these intervals are approximately symmetric around zero, and most point estimates are small. These results show that there are no obvious or notable covariate differences at the cutoff between municipalities where the incumbent party barely won at $t$ and municipalities where the incumbent party barely lost at $t$. 

\subsection*{Outcome Analysis}

\begin{figure}[H]
	\centering
	\caption{Effect of Victory at $t$ on Vote Margin at $t+1$ \\ Incumbent Party, Brazilian Mayoral Elections, 1996-2012 \label{fig:rdplot}}
	\vspace{-0.5in}
	\includegraphics[scale=0.8]{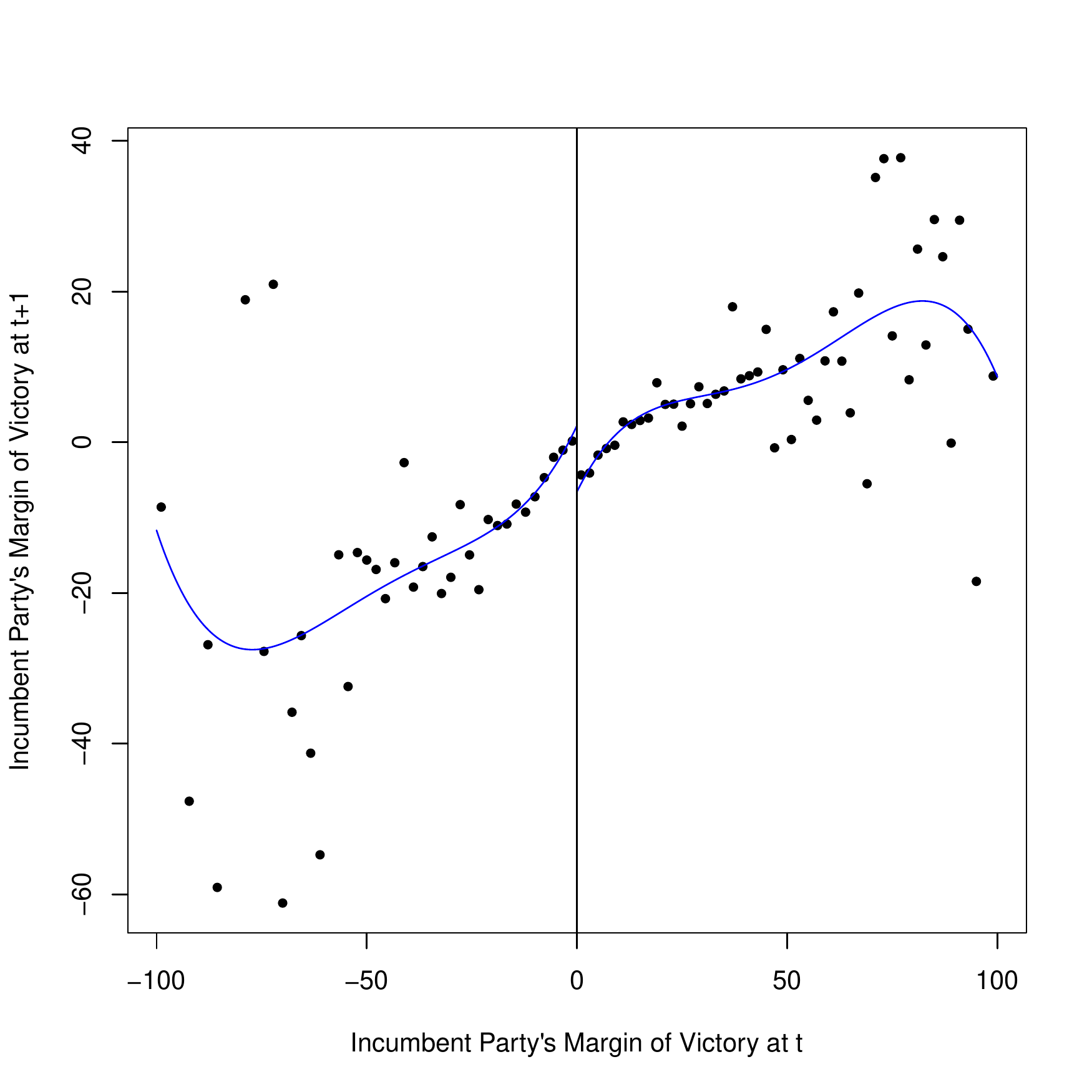}
\end{figure}

Since the evidence from our falsification analysis is consistent with the validity of our RD design, we now proceed to analyze the treatment effect on the main outcome of interest---the incumbent party's margin of victory at $t+1$. This effect is illustrated in Figure \ref{fig:rdplot}. A stark jump can be seen at the cutoff, where the margin of victory of the incumbent party at $t+1$ abruptly decreases as the score crosses the cutoff. This indicates that municipalities where the incumbent party barely wins at $t$ obtain a lower margin of victory at election $t+1$ compared to municipalities where the incumbent party barely loses at $t$, one of the main substantive findings in \citet*{KlasnjaTitiunik2017-APSR}.

We now analyze this effect formally. We first analyze RD effects using the continuity-based framework, employing local polynomial methods with $p=1$ and a MSE-optimal bandwidth. For inference, we use robust bias-corrected 95\% confidence intervals. As we can see in Table \ref{tab:cont}, the MSE-optimal bandwidth is estimated to be around 15.3 percentage points, and within this bandwidth, the RD local-polynomial point estimate is about -6.3. This shows that, at the cutoff, a victory  at $t$ reduces the incumbent party's vote margin at $t+1$ by about 6 percentage points in those municipalities where the party seeks reelection. The 95\% robust bias-corrected confidence interval ranges from -10.224 to -2.945, rejecting the null hypothesis of no effect with a robust p-value of about 0.0004. Including covariates leads to very similar results: the MSE-optimal bandwidth changes to 14.45, and the point estimate moves from -6.28 to -6.10, a very small change, as expected when the covariates are truly predetermined.

\begin{table}[H]
		\caption{Continuity-based RD Analysis: Effect of Victory at $t$ on Vote Margin at $t+1$ \\ Incumbent Party, Brazilian Mayoral Elections, 1996-2012 }\label{tab:cont}
	\begin{tabular}{lcccccc}
		\toprule
                           & RD effect ($\hat{\tau}$)  & 95\% Robust Conf. Int. & Robust p-val & $h$      & $N_l$   & $N_r$   \\
                           \midrule
		Standard           & -6.281 & {[}-10.224 , -2.945 {]} & .0004 & 15.294 & 1533 & 1740 \\
		Using covariates   & -6.106 & {[}-9.881 , -2.656 {]}  & .0007 & 14.453 & 1482 & 1672 \\
		\bottomrule
	\end{tabular}
\end{table}

Second, we analyze the main outcome using a local randomization approach. For this, we must choose the window around the cutoff where the assumption of local randomization appears plausible (if such a window exists). We implement our window selection procedure using the list of covariates mentioned above, an increment of 0.1 percentage points, and a cutoff p-value of 0.15. We use Fisherian randomization-based inference with the difference-in-means as the test statistic and assuming a fixed-margins randomization procedure using the actual number of treated and controls in  each window. As shown in Table \ref{tab:winsel}, starting at the $[0.05, -0.05]$ window and considering all symmetric windows in 0.01 increments, we see that all windows between $[0.05, -0.05]$  and $[0.15, -0.15]$ have a minimum p-value above 0.15. The window $[0.08, -0.08]$ is the first window where the minimum p-value drops below 0.15, indeed, it drops all the way to 0.061. Thus, our selected window is $[-0.15, 0.15]$, which has exactly 38 observations on each side of the cutoff.

\begin{table}[H]
	\caption{Minimum p-value in first 20 symmetric windows around cutoff  \\ Running variable is Vote Margin at $t$ of Incumbent Party, Brazilian Mayoral Elections, 1996-2012 \label{tab:winsel}}
	\centering 
	\begin{tabular}{ccccc}
		\toprule
		Window            & Minimum balance p-val & Covariate of min p-val    &$N_0^-$ & $N_0^+$\\
		\midrule
		{[}0.05,0.05{]} & 0.179           & PSDB previous victory & 10              & 14                  \\
		{[}0.06,0.06{]}   & 0.302         & PSDB previous victory & 13              & 16                  \\
		{[}0.07,0.07{]} & 0.357           & No. effective parties  & 16              & 16                  \\
		{[}0.08,0.08{]}   & 0.231         & No. effective parties  & 18              & 20                  \\
		{[}0.09,0.09{]} & 0.176           & No. effective parties  & 18              & 22                  \\
		{[}0.10,0.10{]}   & 0.34          & PT previous victory & 23              & 28                  \\
		{[}0.11,0.11{]} & 0.335           & Population              & 24              & 30                  \\
		{[}0.12,0.12{]}   & 0.208         & No. effective parties  & 26              & 31                  \\
		{[}0.13,0.13{]} & 0.201           & PT previous victory & 28              & 33                  \\
		{[}0.14,0.14{]}   & 0.167         & No. effective parties  & 34              & 36                  \\
		{[}0.15,0.15{]} & 0.157           & No. effective parties  & 38              & 38                  \\
		{[}0.16,0.16{]}   & 0.062         & PT previous victory & 42              & 41                  \\
		{[}0.17,0.17{]} & 0.114           & PT previous victory & 43              & 43                  \\			{[}0.18,0.18{]}   & 0.044         & PT previous victory & 49              & 45                  \\
		{[}0.19,0.19{]} & 0.065           & PT previous victory & 51              & 50                  \\
		{[}0.20,0.20{]}     & 0.054       & PT previous victory & 53              & 50                  \\
		\bottomrule
	\end{tabular}
\end{table}

\begin{figure}[H]
	\centering
	\caption{Window Selector Based on Covariates \\ Incumbent Party, Brazilian Mayoral Elections, 1996-2012 \\ Running variable is Incumbent party's Margin of Victory at $t$ }	\label{fig:pvalues}
	\includegraphics[scale=0.9]{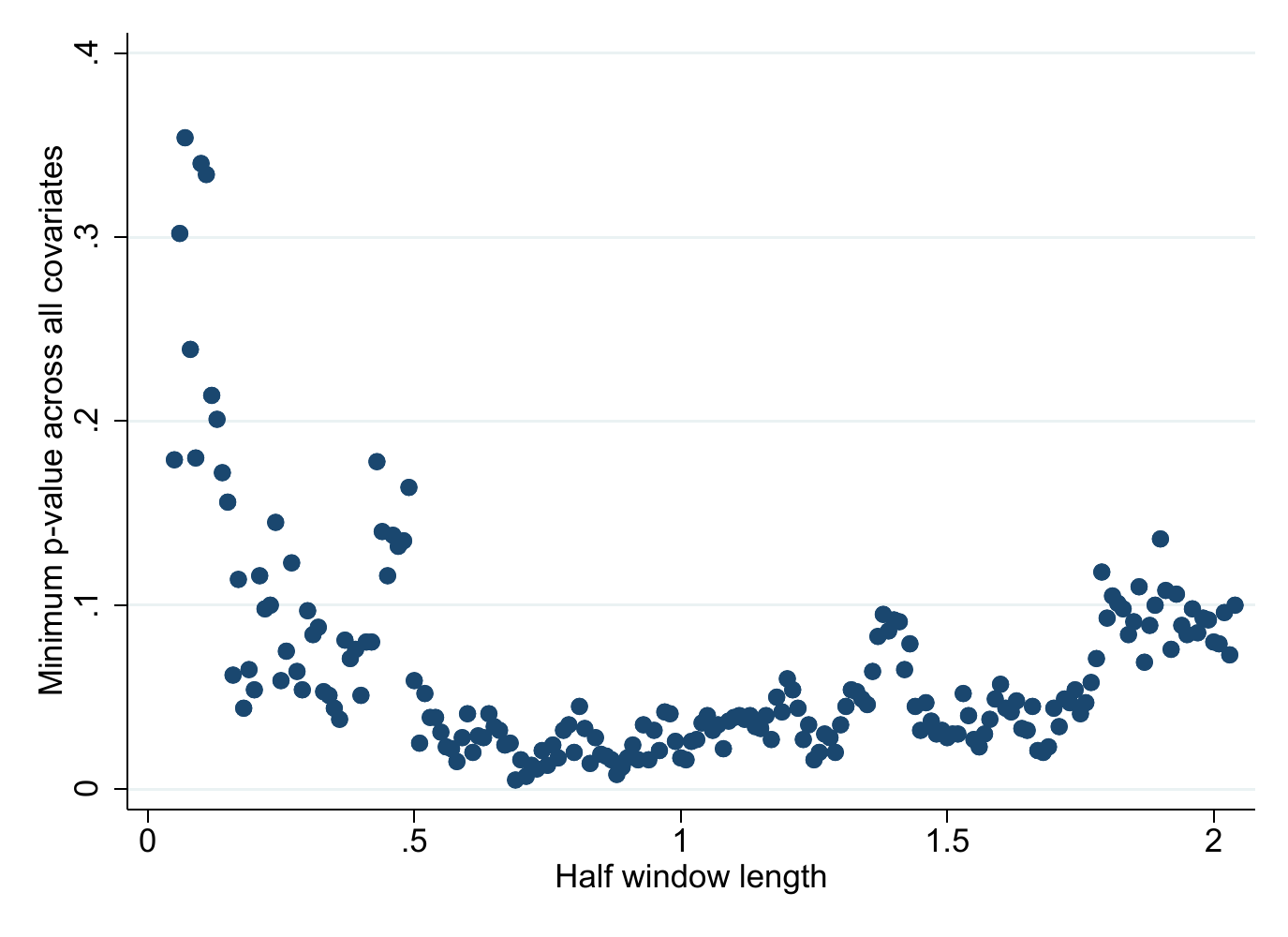}
\end{figure}

In order to further illustrate the results in Table \ref{tab:winsel}, Figure \ref{fig:pvalues} shows the associated p-values for all  symmetric windows in 0.01 increments between $[0.05, -0.05]$  and $[-2.00, 2.00]$.

In table \ref{tab:locran}, we present our inference results in the chosen window $[-0.15, 0.15]$, reporting both Fisherian inference (using the same choices as those used in the window selection procedure) and large-sample p-values. The treated-control difference-in-means is $-9.992$, with a Fisherian p-value of approximately $0.083$ and a large-sample p-value of about $0.070$, rejecting both the sharp null hypothesis and the hypothesis of no average effect at 10\% level. The fact that the point estimate continues to be negative and that the p-values are 8\% and below suggests that the continuity-based results are broadly robust to a local-randomization assumption, as both approaches lead to similar conclusions. The local randomization p-value is much larger than the p-value from the continuity-based local polynomial analysis, but this is likely due, at least in part, to the loss of observations, as the sample size goes from a total of 3,412 (1,740+1,672) observations to just 39 (19+20). (The discrepancy in the number of observations in $[-0.15,0.15]$  between the outcome analysis and the window-selector analysis stems from missing values in the covariates.)

\begin{table}[H]
	\caption{Local Randomization RD Analysis: Effect of Victory at $t$ on Vote Margin at $t+1$ \\ Incumbent Party, Brazilian Mayoral Elections, 1996-2012 \label{tab:locran}}
	\centering
	\begin{tabular}{cccccccc}
		\toprule
		& RD effect $\hat{\tau}_0$           & Fisher p-val  & Large-sample p-val  & Window  &$N_0^-$ & $N_0^+$\\
		\midrule 
		& -9.992 & 0.083   & 0.0697 & [-0.15, 0.15] &    19 & 20\\
		\bottomrule   
	\end{tabular}
\end{table}

\clearpage
\section{Final Remarks}

We reviewed two alternative frameworks for analyzing sharp RD designs. First, the continuity-based approach, which is more common in empirical work, assumes that the unknown regression functions are continuous at the cutoff. Estimation is conducted nonparametrically using local polynomial methods, and bandwidth selection relies on minimizing a criterion such as the MSE or the coverage error probability. Inference under this framework relies on large sample distributional approximations, and requires robust bias correction to account for misspecification errors local to the cutoff. Second, the local randomization approach formalizes the intuition that RD designs can be interpreted as local experiments in a window around the cutoff. In this case, the window is chosen to ensure that treated and controls are comparable in terms of observed predetermined characteristics, as in a randomized experiment. Within this window, inference is conducted using randomization inference methods assuming that potential outcomes are non-random, or other canonical analysis of experiments methods based on large sample approximations.

These two approaches rely on different assumptions, each with its own advantages and disadvantages, and thus we see them as complementary. On the one hand, the continuity-based approach is agnostic about the data generating process and does not require any modeling or distributional assumptions on the regression functions. This generality comes at the expense of basing inference on large-sample approximations, which may not be reliable when the sample size is small (a case that is common in RD designs, given their local nature). On the other hand, the Fisherian local randomization approach provides tools to conduct inference that is exact in finite samples and does not rely on distributional approximations. This type of inference is more reliable than large-sample-based inference when the sample size is small. And if the sample size near the cutoff is large, the analysis can also be conducted using standard large-sample methods for the analysis of experiments. However, the conclusions drawn under the local randomization approach (either Fisherian or large-sample) require stronger assumptions (unconfounded assignment, exclusion restriction) than the continuity-based approach, are conditional on a specific sample and window, and do not generalize to other samples or populations.

In sum, as in \citet*{Cattaneo-Titiunik-VazquezBare_2017_JPAM}, we recommend the continuity-based approach as the default approach for analysis, since it does not require parametric modeling assumptions and automatically accounts for misspecification bias in the regression functions when conducting estimation and inference. The local randomization approach can be used as a robustness check, especially when the sample size is small and the large-sample approximations may not be reliable. 

There is one particular case, however, in which the continuity-based approach is not applicable: when the running variable exhibits only a few distinct values or mass points (even if the sample size is large because of repeated values). In this case, the nonparametric methods for estimation, inference, and bandwidth selection described above do not apply, since they are developed under the assumption of local approximations and continuity of the score variable, which are  violated by construction when the running variable is discrete with a small number of mass points. Thus, in settings where the running variable has few mass points, local randomization methods, possibly employing only the closest observations to the cutoff, are a more natural approach for analysis. We refer the reader to \citet*{Cattaneo-Idrobo-Titiunik_2019_Book2} for a more detailed discussion and practical illustration of this point.

\newpage
\bibliographystyle{jasa}
\bibliography{Cattaneo-Titiunik-VazquezBare_2020_Sage}

\newpage
\appendix

\end{document}